# DUAL BAND GNSS ANTENNA PHASE CENTER CHARACTERIZATION FOR AUTOMOTIVE APPLICATIONS

Ran Liu and Daniel N. Aloi

Electrical and Computer Engineering Department,
Oakland University, Rochester, MI, USA

## ABSTRACT

*High-accuracy Global Navigation Satellite System (GNSS) positioning is a prospective technology that will be used in future automotive navigation systems. This system will be a composite of the United States' Global Positioning System (GPS), the Russian Federation's Global Orbiting Navigation Satellite System (GLONASS), China Beidou Navigation Satellite System (BDS) and the European Union's Galileo. The major improvement in accuracy and precision is based on (1) multiband signal transmitting, (2) carrier phase correction, (3) Real Time Kinematic (RTK). Due to the size and high-cost of today's survey-grade antenna solutions, this kind of technology is difficult to use widely in the automotive sector. In this paper, a low-cost small size dual-band ceramic GNSS patch antenna is presented from design to real sample. A further study of this patch antenna illustrates the absolute phase center variation measured in an indoor range to achieve a received signal phase error correction. In addition, this low-cost antenna solution is investigated when integrated into a standard multi-band automotive antenna product. This product is evaluated both on its own in an indoor range and on a typical vehicle roof at an outdoor range. By using this evaluation file to estimate the receiver position could achieve phase motion error-free result.*

## KEYWORDS

*Antenna Phase Determination, Real-Time Kinematic (RTK), Autonomous Drive, Phase Center Variation (PCV) and Phase Center Offset (PCO).*

## 1. INTRODUCTION

Global Navigation Satellite Systems (GNSS) have been widely developed in the last 30 years. A civilian user's position accuracy has improved during that time from ±5 meters to ±1 meter. To fulfill the requirement of upcoming autonomous vehicles, a ±1m positional accuracy will not be sufficient. There are several potential ways to improve the position accuracy that could be utilized in a low-cost mass production vehicle, such as multi-system aliasing, multi-band signal utilization and carrier phase correction. Conceptual calibration of antenna phase characteristic was presented in the 1990s, including absolute method [1] and relative method [2]. In the past few decades, anechoic chamber of the absolute method antenna phase information had been studied with high precise GNSS technology [3-4]. Recently there is a study work on the phase center variations for GNSS signal processing based on the chamber measurement [5]. The above studies are based on professional survey grade antenna in the chamber, which are not feasible for mass civil application. After that a comparison work of characterizing antenna phase center with high-cost and low-cost antenna is presented in [6].

In this study, the work is focused on the low-cost GNSS antenna performance of a vehicular platform, especially for the upcoming autonomous driving vehicle. This paper emphasizes multi band signal utilization and carrier phase correction, specifically a dual-band L1/L5 GNSS antenna design with carrier phase correction via Phase Center Variation (PCV) chamber measurement and Phase Center Offset (PCO) evaluation. The main contribution of this paper was the characterization of antenna phase center motion error on the PCV and PCO for a dual-band GNSS patch antenna: 1) on a 250mm circular ground plane; 2) integrated into a shark-fin antenna on this ground plane; and 3) integrated into a shark-fin antenna on a vehicle roof. Each scenario has different PCV and PCO performance. This means that each vehicle platform and antenna location require a unique calibration to characterize PCV and PCO for optimal position accuracy estimation.

This paper is organized as follows; firstly, understanding the antenna phase center motion error in a positioning system, then the multi-band GNSS strategies of different OEMs, after that an applicable design for L1/L5 antenna patch and its phase result, finally some of measurement results of the proposed design from both indoor and outdoor antenna ranges.

## 2. HIGH ACCURACY AND PRECISION POSITIONING

The principle of satellite navigation is the satellite-user range measurements based on the times of transmission and receipt of signal, which includes two major methods: code pseudorange Equation (1) and carrier phase Equation (2) measurements. The carrier phase measurement adds the carrier wave and integer ambiguity $N$, the rest of equation part is almost the same as the code pseudorange. The reason the carrier phase could significantly improve the position accuracy and precision is that measurement resolution is much finer than the code measurement. The L1 C/A code chipping rate is 1.023 MHz, the corresponding wavelength is around 300 m, while the carrier phase of L1 is 1575.42 MHz, and the corresponding wavelength is around 0.19 m. By analogy with the wavelength as a measuring tape, carrier phase measurement has a fine tick mark. On the other hand, the code pseudorange period is known as a digital signal, thus $\rho$ is an easy acquisition; while the carrier phase φ has difficulty finding the transmission wave period of integer ambiguity $N$ and system error $\varepsilon_{\varphi}$.

$$\rho = d + I_{\rho} + T_{\rho} + c[\delta t_u - \delta t^s] + \varepsilon_{\rho} \tag{1}$$

$$\varphi = \frac{1}{\lambda}\left[d + I_{\varphi} + T_{\varphi}\right] + \frac{c}{\lambda}(\delta t_u - \delta t^s) + N + \varepsilon_{\varphi} \tag{2}$$

Where $\rho$ is the code pseudorange, $d$ is the true range distance between the user and satellite, $I_{\rho}$ is ionospheric pseudorange delay, $T_{\rho}$ is tropospheric delay, $c$ is the speed of the light in a vacuum; $\delta t_u$ and $\delta t^s$ are denoted as user receiver and satellite clock biases (unit: seconds), $\varepsilon_{\rho}$ is pseudorange error includes unmodeled effects, modeling error, and measurement errors;

φ is the transmission period phase of satellite to user signal, $\lambda$ is transmission signal wavelength (unit: meters), $I_{\varphi}$ is carrier phase advance, while the phase velocity of carrier in the ionosphere exceeds the light speed in a vacuum, $I_{\rho}$ and $I_{\varphi}$ are equal in magnitude but opposite in sign ($I_{\rho} = -I_{\varphi}$), $T_{\varphi}$ is tropospheric delay, $N$ is the period integer ambiguity, $\varepsilon_{\varphi}$ is carrier phase error (3)

$$\varepsilon_{\varphi} = \varepsilon_{phase} + \varepsilon_{unmodel} + \varepsilon_{modeling} + \varepsilon_{meas} \tag{3}$$

The term carrier phase measurement error $\varepsilon_{\varphi}$ includes $\varepsilon_{phase}$ antenna phase-center motion error, $\varepsilon_{unmodel}$ unmodeled effects, $\varepsilon_{modeling}$ modeling error, and $\varepsilon_{meas}$ measurement errors. $\varepsilon_{phase}$ antenna phase-center motion error includes.

$$\varepsilon_{phase} = \varepsilon_{r,PCV} + \varepsilon_{r,PCO} + \varepsilon_{PCV}^{s} + \varepsilon_{PCO}^{s} + \varepsilon_{disp} + \varepsilon_{pw} \quad (4)$$

$\varepsilon_{r,PCV}$ and $\varepsilon_{r,PCO}$ are receiver antenna phase center variation and offset; $\varepsilon_{PCV}^{s}$ and $\varepsilon_{PCO}^{s}$ are satellite antenna phase center variation and offset. $\varepsilon_{disp}$ is site displacement and $\varepsilon_{pw}$ is phase wind-up effect [8]. $\varepsilon_{phase}$ can vary with the direction of arrival (azimuth and elevation) of a signal and such variation can range under centimeters or even millimeters. The main purpose of this paper is to determine $\varepsilon_{r,PCV}$, $\varepsilon_{r,PCO}$ , and evaluate these two errors and their impact on high precision positioning.

Real-Time Kinematic (RTK) is a technology could achieve high accuracy and precision positioning, which could resolve the above (2) unknowns, but not receiver antenna phase center motion error $\varepsilon_{r,PCV}$ and $\varepsilon_{r,PCO}$ in $\varepsilon_{\varphi}$; by using the RTK methodology

1) Single difference eliminates the satellite clock biases $\delta t^{(k)}$, ionospheric delay $I$ and tropospheric delay $T$;
2) Double difference eliminates clock bias in reference $\delta t_r$ and user $\delta t_u$;
3) Triple difference eliminates the integer ambiguity $N$.

## 3. Multi Frequency Band Benefit

In order to make a simple expression, the word "L1" in this paper will represent GPS Link 1 (1574.42-1576.42 MHz), Glonass G1 (1593-1607 MHz) and Beidou B1 (1559-1561 MHz) , "L2" is GPS Link 2 band (1215-1237 MHz) and "L5" is GPS Link 5 band (1164-1189 MHz).

In the range fundamental code pseudorange (1) and carrier phase (2) measurement, the ionospheric delay $I$ can contribute a large portion of the positional error. This is especially true due to slant delay, when the vertical component of a circularly polarized electromagnetic wave is influenced by the angle of incidence through the ionosphere. The diversity of signal frequencies could significantly overcome the ionospheric delay as well as reduce tracking noise in multipath propagation [7]. Although ionospheric delay could be eliminated in a condition by using the single difference method, another approach is a dual-frequency (ex. L1 and L5) receiver can estimate the $I_{\rho}$ and $I_{\varphi}$ from the measurement without a reference base station. By the physics discovery, ionized gas is a dispersive medium for radio waves. The refractive index depends on a radio wave frequency [7].

The estimate of pseudorange measurement ionospheric group delay at L1 is

$$I_{\rho}(L1) = \frac{f_{L5}^{2}}{f_{L1}^{2} - f_{L5}^{2}}(\rho_{L5} - \rho_{L1}) \quad (5a)$$

The estimate of carrier phase measurement ionospheric phase advance at L1 is

$$I_{\varphi}(L1) = -\frac{f_{L5}^{2}}{f_{L1}^{2} - f_{L5}^{2}}[\lambda_{L1}(\varphi_{L1} - N_{L1}) - \lambda_{L5}(\varphi_{L5} - N_{L5})] \quad (5b)$$

The ionospheric error of $I_\rho$ and $I_\varphi$ can be both estimated, however $I_\rho$ $(5a)$ is unambiguous, while $I_\varphi(5b)$ involves integer ambiguities $N$. A practiced way to solve integer ambiguities is using code measurements to estimate integers [7].

Based on this idea, the dual-band and even triple-band GNSS receiver are well known and implemented in the automotive industry. In the next section, a multi-band GNSS antenna system will be discussed in detail.

## 4. ANTENNA SYSTEM DIAGRAM

As mentioned in the previous section, the $\varepsilon_{r,PCV}$ and $\varepsilon_{r,PCO}$ in Equation (4)are effects from the receiver's antenna. A decent antenna system design is necessary. A multi-band measurement has an improvement in the estimation of the ionospheric delay, as a result, automotive OEMs are pushing multi-band receiver implementation with an emphasis on the antenna's performance that now includes phase variation.

From the Table 1 shown below, the multi-band GNSS antenna strategy varies in different OEMs. OEM1 and OEM2 are American based companies, while OEM3 (with additional L-Band correction service 1525-1550 MHz) and OEM4 are European based manufacturers. After an in-depth discussion with different OEMs, the L1/L5 multi-band is selected for this research case. The reasons for L5 instead of L2 are the following:

- The L5 band provides a 10.23 MHz chipping rate, which is the fastest chipping rate available in civilian code.
- L5 band is intended to meet the needs of critical safety-of-life applications
- For a single antenna, it is hard to cover the L2 and L5 frequency range simultaneously due to the wide bandwidth (1164-1237 MHz).

Table 1. OEMs Multi Band GNSS Strategy

| | L1 (GNSS) 1559-1607 MHz | L2(GPS) 1215-1237 MHz | L5(GPS) 1164-1189 MHz |
|---|---|---|---|
| OEM1 | Yes | | Yes |
| OEM2 | Yes | | Yes |
| OEM3 | Yes | Yes | |
| OEM4 | Yes | Yes | Yes |

A common antenna system is composed of two parts, as shown in Figure 1, the passive antenna and the active LNA circuit. This paper is mainly focused on the passive section's design and evaluation, including the antenna design and ceramic patch selection with consideration towards gain, axial ratio (AR), and the last but not the least, PCV and PCO. The PCV and PCO performance results will be used to evaluate the antenna's performance in three stages; firstly on a circular 250 mm diameter ground plane, secondly with the patch integrated into a compact shark-fin product on the same ground plane and lastly with that same shark-fin product mounted on a vehicle roof.

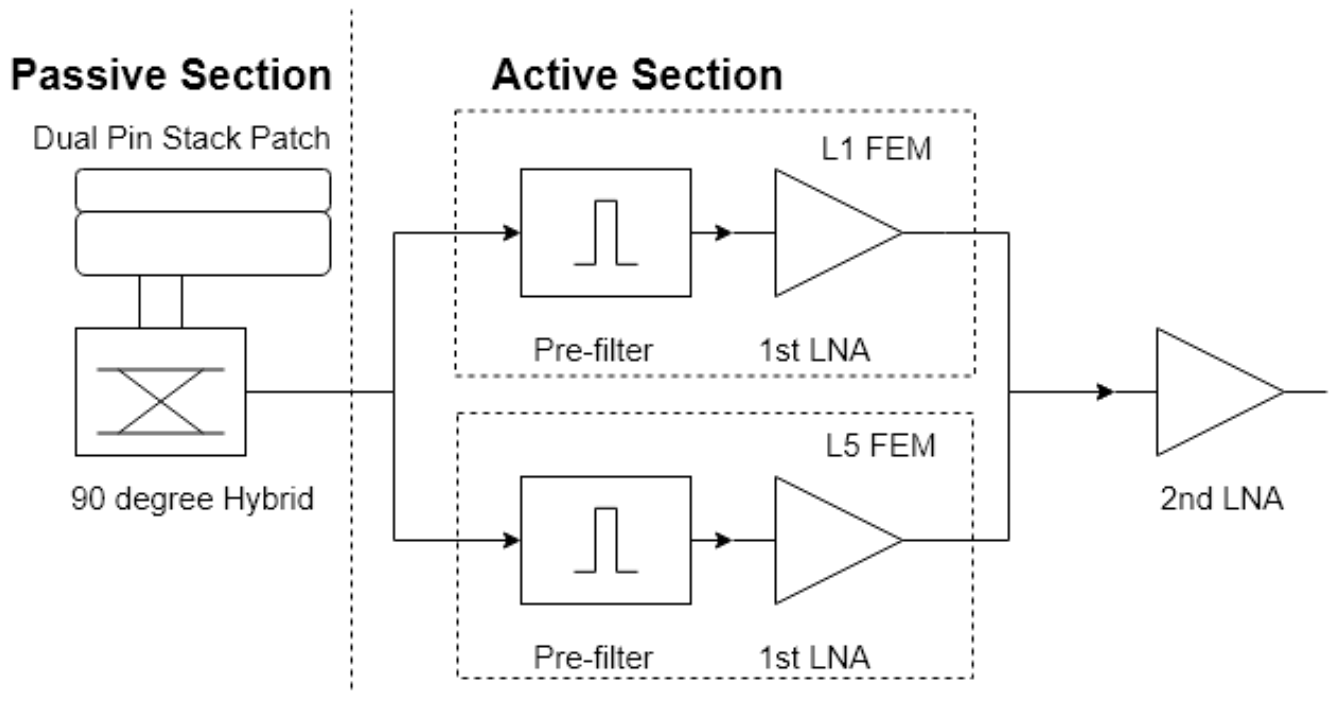

Fig. 1. Antenna System Diagram

## 5. PASSIVE SECTION DESIGN

Antenna selection and design is the first important step in designing the system. Passive antenna performance will be determined by the receiving system's gain, which is the signal strength on pseudorange $\rho$ from Equation $(1)$. Meanwhile the antenna performance has an impact on Dilution of Precision (DOP), which specifies the effect that navigation satellite geometry has on positional measurement precision. In designing a product for the automotive industry, the cost control and mass production quality are both important topics not to be avoided. Thus, a good balance of antenna performance and price needs to drive the design process.

The passive antenna is the first element to receive the signal from the satellite and will therefore significantly influences the system performance, so the type of antenna needs to be carefully chosen. Table 2 provides an overview of some potential antenna options common in the automotive industry.

Table 2. Potential GNSS Antennas Options

| | Material | Size | Production | Price |
|---|---|---|---|---|
| Ceramic Patch | Ceramic (DK=10/13/20) | Small | Easy | Medium |
| Spiral | Teflon (DK=2.2) | Medium | Hard | Medium |
| 3D- Ring | Air (DK=1) | Big | Hard | Cheap |
| Scarabaeus[9] | Air | Big | Hard | Cheap |

*DK: dielectric constant

From the Table 2, the ceramic patch with its small package size and simple method of mass production assembly appears to be a good starting choice given the mass production, reliability and cost requirements. With consideration towards the strict PCV and PCO requirements, square patches generally have a good symmetric surface current distribution, and thus it may have a better PCV. Therefore, this study focuses on a square ceramic patch antenna design in the following sections.

In order to find the patch antenna field radiated performance, it can be evaluated magnetic current density ($\boldsymbol{M_1}$ $and$ $\boldsymbol{M_2}$) and electric field ($\boldsymbol{E_1}$ $and$ $\boldsymbol{E_2}$) by patch slots, as slot1 and slot 2 examples are demonstrated in Figure 2 $(a)$. The theoretical calculation is based on the perfect boundary condition. On the following section, the field result could get influence by the boundary condition changes. Furthermore, the later antenna far-field phase result will explain the degradation.

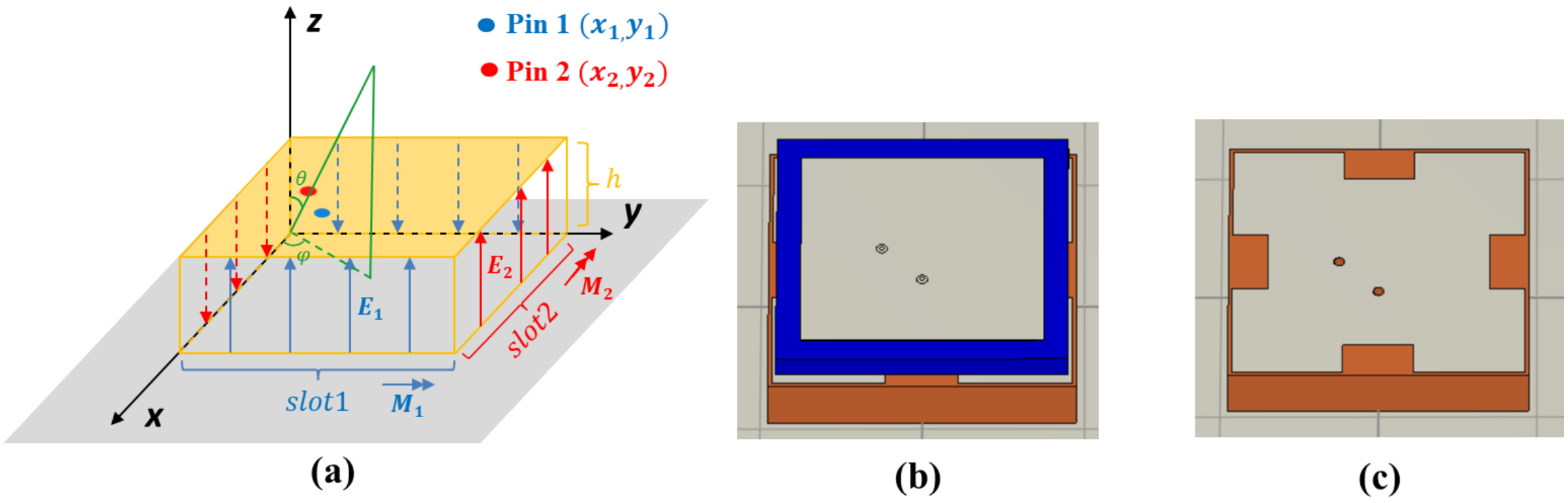

Fig. 2. (a) Rectangular 2-pin Patch with Radiating Slots and Equivalent Magnetic Current Densities; The patch Antenna simulation model
(b) Top Layer for L1 band; 36x36x3mm (LWH), DK 10;
(c) Top Layer for L5 band, 38x38x7mm (LWH), DK 13.

The perfect boundary condition:
Electrical Wall: patch top and bottom

$$\boldsymbol{E_x}(z' = 0 \; and \; z' = h) = 0 \quad (6a)$$
$$\boldsymbol{E_y}(z' = 0 \; and \; z' = h) = 0 \quad (6b)$$

Magnetic Wall: from top to bottom patch

$$\boldsymbol{H_z}(0 < z < h) = 0 \quad (6c)$$

The antenna is fed into two orthogonal out of phase pins ($TM^z_{100}$ and $TM^z_{010}$ modes) on the top layer, which is a pair of orthogonal degenerate modes. Out of phase port, is generated by a 90-degree hybrid coupler (Figure 1), right-hand polarization in this case.

An open-circuit electric field is on the gap between feeding pin and L5 band patch. This open-circuit electric field is equivalent as “magnetic element”, which could excite L5 band frequency. In general, the top and bottom patches with different resonate dimension could excite dual frequency simultaneously by a single pin [10-11].

The dual-band GNSS system can utilize multiple frequency bands to reduce the major ionospheric error in Equation $(5a)$ and $(5b)$. The challenge is to design an antenna system that can efficiently and affordably mitigate ionospheric error in order to increase GNSS positional accuracy. Considering each frequency phase center and the receiver module architecture, a stacked patch will have less phase center variation and cable bias than two separate patches. The above discussion is a general antenna selection guideline and it may have other applicable approaches.

A right hand circularly polarized (RHCP) ceramic patch antenna is commonly used in modern industry products, because of its physical stability, low profile, manufacturability and affordable cost. A typical patch antenna is designed and simulated by a commercial off-the-shelf (COTS) three-dimensional, full wave electromagnetics (EM) field Time Domain solverof CST based off the method of time domain finite integration technique (FIT). As shown in Figure 2 $(b)$, the top layer (blue) is designed to receive the L1 band signal. The top L1 stacked patch $(LWH: 36 \times 36 \times 3mm)$ material has a ceramic dielectric constant (DK) 10. The bottom layer (orange) is designed to receive the L5 band signal, as shown in Figure 2 $(c)$, the bottom L5 stacked patch $(LWH: 38 \times 38 \times 7mm)$ material is DK 13, and the total height after assembly including adhesive tape is $11mm$. In order to fulfill vehicle styling, a compact shark-fin antenna design is widely implemented. Thus, the patch antenna's dimensions are limited by the shark-fin antenna housing.

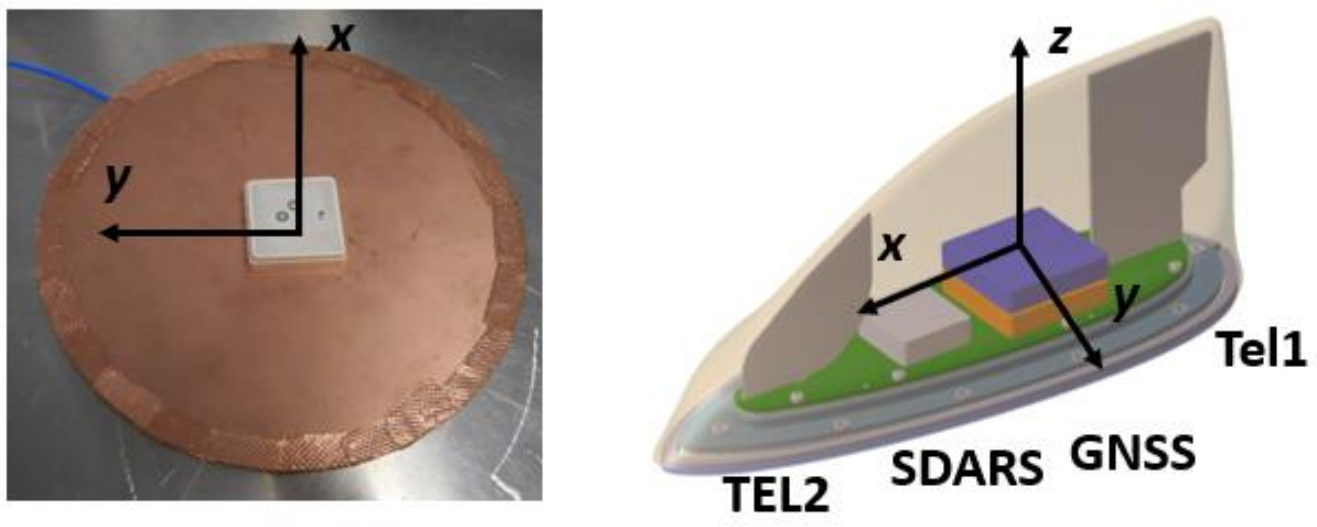

(a) (b)

Fig. 3. (a) Real Patch EVB on 250mm Circle GND (b) Isometric overview of shark-fin antenna on 250mm Circle GND

A few engineering samples are fabricated to verify the design and these samples are mounted on a ground plane. Figure 3 $(a)$ is a standalone real patch antenna mounted on an evaluation board (EVB), which is on a 250 mm circular ground plane. Figure 3 $(b)$ illustrates a production intent shark-fin antenna with an isometric view in CAD software. The proposed L1/L5 dual-band GNSS antenna is highlighted with the same coloring scheme of Figure 2. The shark-fin radome also houses a satellite digital audio radio services (SDARS) patch antenna located in the front of GNSS patch and two cellular radiators (Tel1 and Tel2) at rear/front side of the shark-fin antenna. A real shark-fin antenna is mounted on a 250 mm circular ground plane as well for comparison testing at the same time.

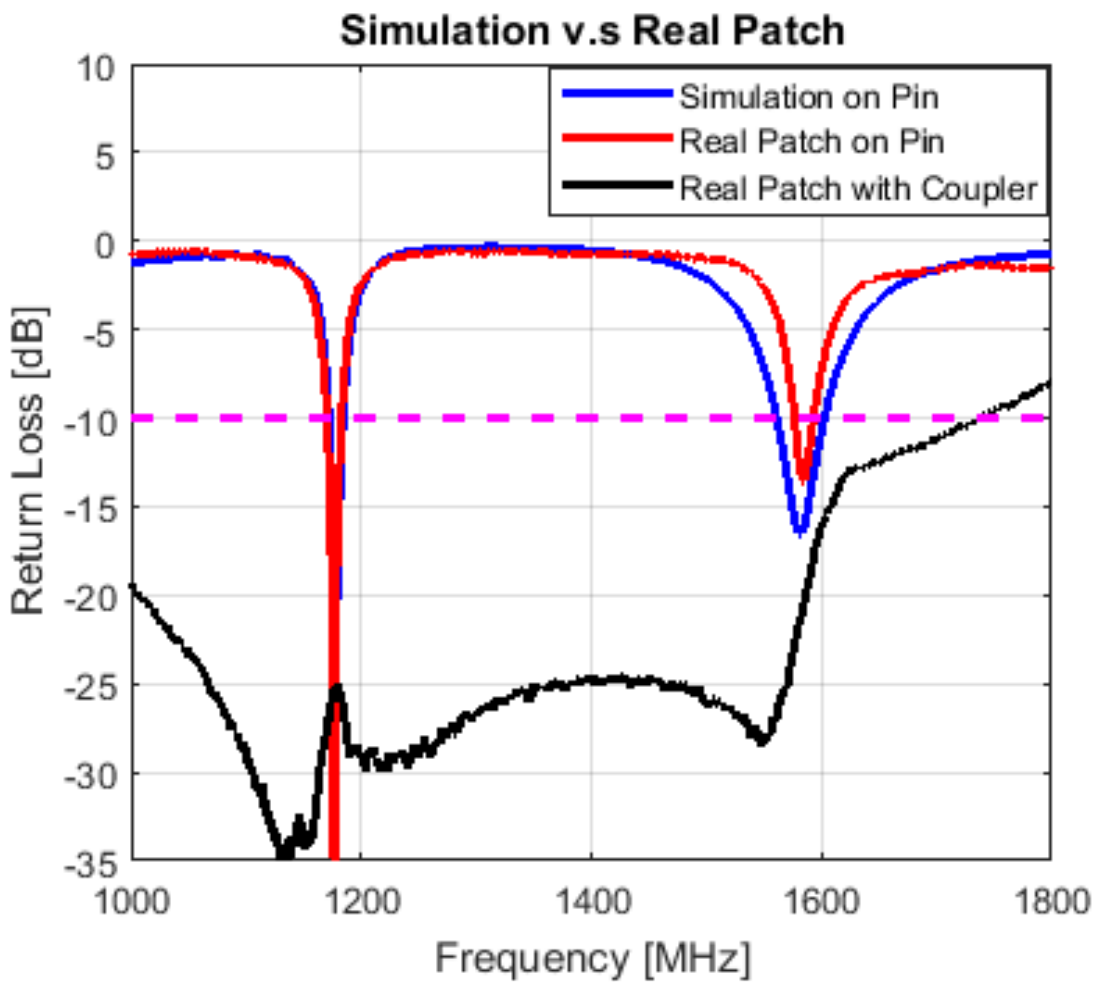

Fig. 4. Return loss result of simulation and real patch

This solution is acceptable due to the fairly narrow desired range for the L5 frequency band (1167-1189 MHz) and L1 frequency band (1559-1607MHz), the result shown in Figure 4. From the comparison result between simulation model (blue) and a real patch antenna (red) at feeding pin, the return loss curves are almost matched. The antenna passive section (Figure 1) is including the patch antenna and 90-degree hybrid coupler, the measurement result at coupler output port (black) have a wide band matching performance to cover L1 and L5. In the practice, return loss bandwidth should less than -10 dB.

The evaluation board simulation result of surface current densities is demonstrated in Figure 5. In this quasi perfect boundary condition, the most current densities have symmetric distribution over the patch center line (*y-axis*). When the port excited at L5 1176MHz with 90-degree phase signal, the maximum current densities is around 39.27 dB (A/m) underneath of the patch's ground plane. On the patch two-side edge's ground plane left and right, the red arrow shown the surface current densities achieve up to 24.26 dB (A/m), then it gets the degradation by the distance to the edge.

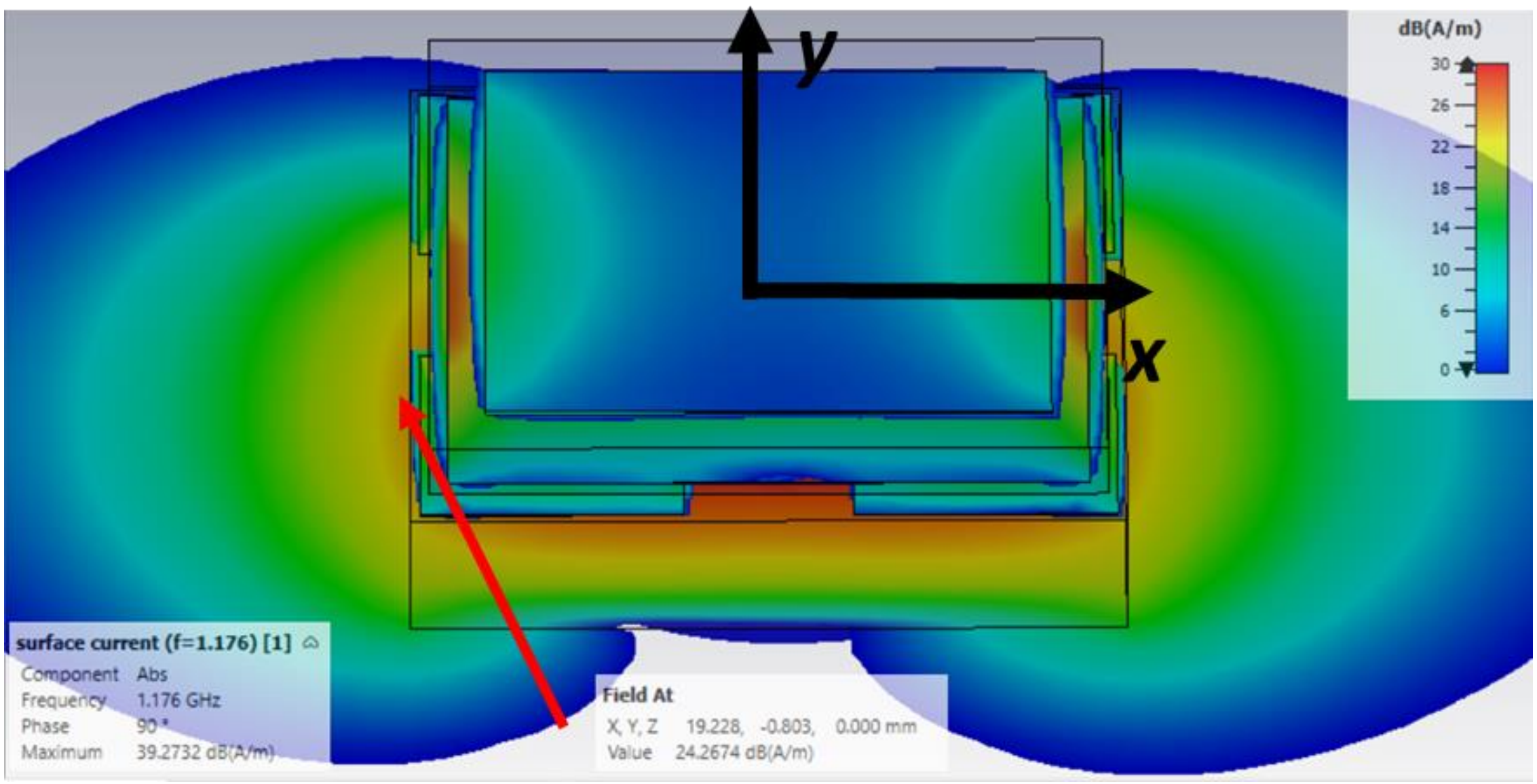

Fig. 5. Simulation result of surface current densities @ EVB 1176MHz

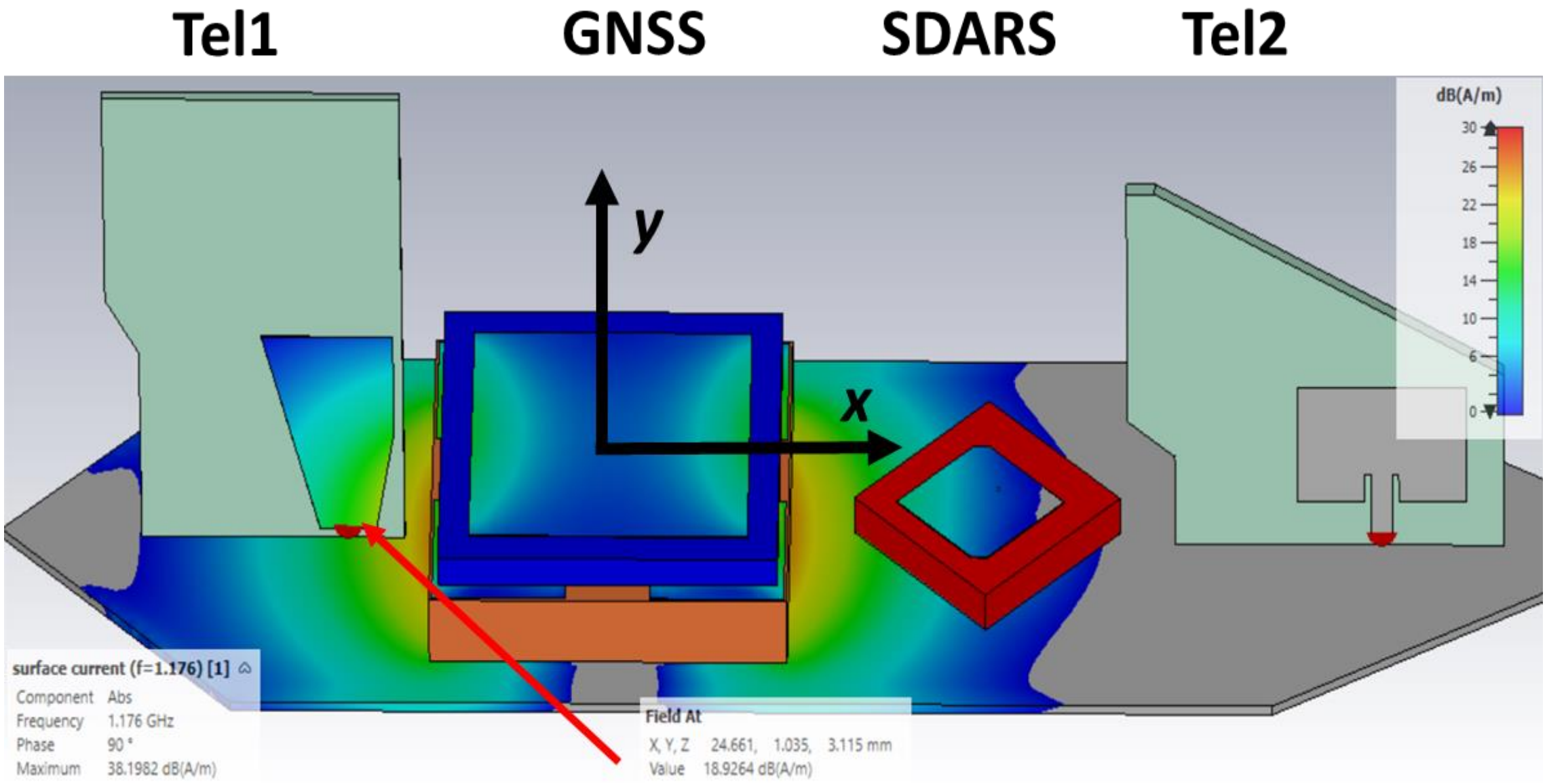

Fig. 6. Simulation result of surface current densities @ Sharkfin 1176MHz

In the meantime, a shark-fin antenna lite CST simulation model without delicacy mechanical structure is built up in Figure. 6. As mention above, the small ground size of the GNSS antenna would limit the electrical wall condition Equation $(6a\ and\ 6b)$,the antenna surrounding elements would break the GNSS patch antenna magnetic wall boundary condition Equation $(6c)$, due the shark-fin antenna compact packaging. As the simulation result shown the Tel1 element and SDARS patch have the influence on the surface current densities. As the marker shown on Tel1 bottom point, the coupling current densities could achieve up to 18.92 dB (A/ m). Due to the height of the Tel1 element, there is a coupling current distribution on the vertical direction. This kind of impact would involve the GNSS antenna performance at low elevation angle until the patch antenna have an open sky view. The similar coupling current densities on the SDARS patch antenna surface, it would be around 11.78 dB (A/ m).

On L1 frequency result, the simulation shown the similar behavior. While the L1 1575.42 MHz is more closed to the cellular Band 3 (1710-1880 MHz), the Tel1 element would have more impact on the L1 band than the L5 band. The real measurement result will be shown on the coming Section VII.

Comparing with Figure 5 and 6, they show the GNSS antenna nearby element would have an isolation problem based on the surrounding surface current densities analysis, the coupling current on the surrounding element would break the GNSS antenna boundary condition, further degradation involves the antenna far-field performance. On the design stage, there are several ways to reduce the influence, such as

- A certain distance between the elements, Tel1 against with GNSS patch's distance; SDARS patch with GNSS patch's distance;
- Telephone antenna resonate structure with GNSS frequency notch out;
- SDARS patch rotation in certain angle.
- Anti-resonate frequency tuning away GNSS in-band frequency.

The methods are not limited in above all, it may have different approaches. From the EM software of E-field/H-field/current densities simulation result, it provides a visual method to guide design work. It is also important to have a bench sample in hand-made tuning. However, the co-existing problem has a mutual impact, the surrounding elements would get performance degradation as well within the GNSS antenna nearby. The golden rule of a better decoupling solution is having further physical distance.

## 6. INDOOR ANTENNA RANGE SETUP

An indoor far-field antenna range is provided by Continental Advanced Antenna Inc, USA. The distance between transmission antenna and antenna under test (AUT) is 9.14 meters, which meets the far-field requirement in both L1 and L5 frequency bands. The antenna test range's setup is demonstrated in Figure 7 $(a)$. The indoor test range is set for a spherical coordinate system. The configuration of the upper hemisphere is such that the angle of antenna elevation (90 - $\theta$): 0 to 90°, with a step of 1° and the angle of antenna azimuth ($\varphi$): 0 to 360°, with a step of 5°. The patch antenna is mounted on an offset arm, which ensures the AUT is in the rotation center (Figure 7 $(b)$). In this setup, the initial rotation center is called the antenna reference point (ARP) $\theta$ = 0° or elevation = 90°, $\varphi$=0°, PCV=0 mm.

The vertical electrical plane radiation pattern and horizontal electrical plane radiation pattern are generated by the two-port cross-dipole parabolic reflector antenna (Figure 7 $(c)$). The incident signal transmits through the vertical (V) and horizontal (H) port cables to vector network analyzer. After post- processing, the output phase delay depends on the "chirality" of polarization. A right-hand circular polarization (RHCP) and left-hand circular polarization (LHCP) are analyzed in the following section. A narrow beam (HPBW = 18°) is excited by this directional parabolic reflector antenna, which reduces the radiation reflection from the walls inside the anechoic chamber.

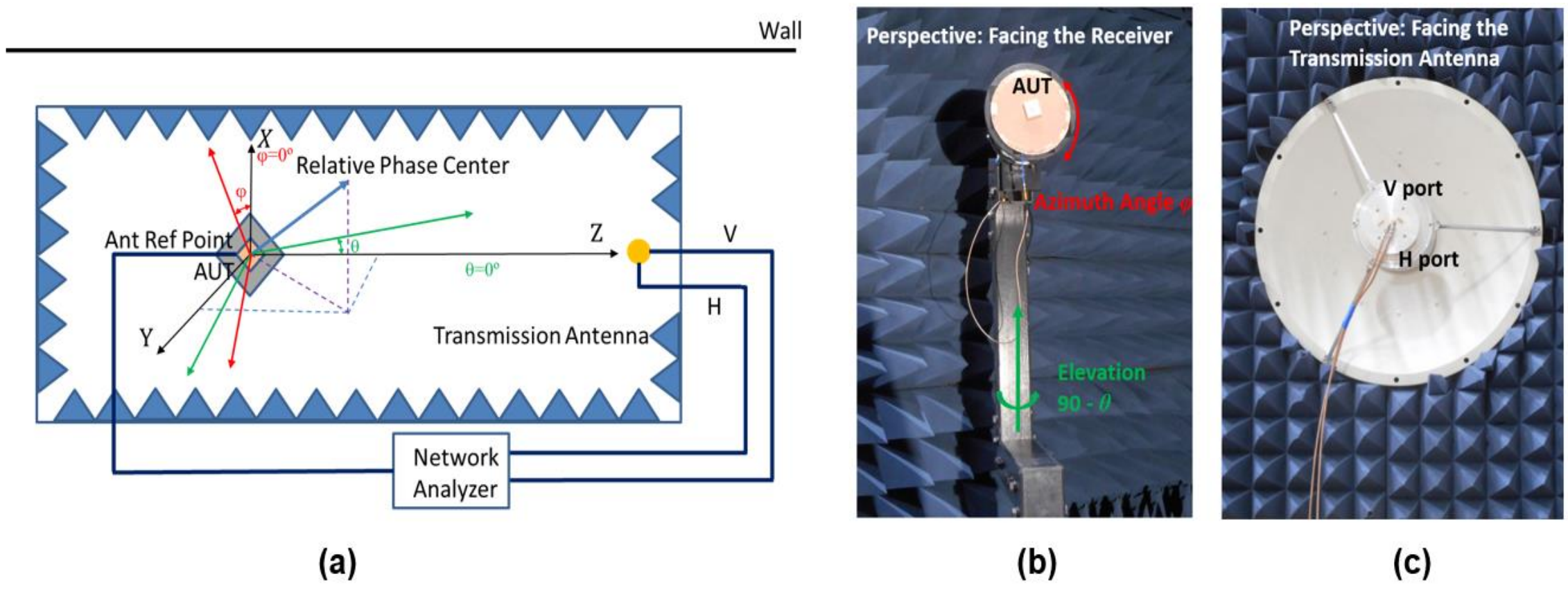

Fig. 7. Indoor Range Setup; (a) Diagrammatic Drawing; (b) AUT and Turntable; (c) Transmission Antenna

## 7. ANTENNA GAIN AND AXIAL RATIO

Traditionally, GNSS antenna performance is evaluated by the antenna's linear average gain (LAG) and axial ratio (AR) over the elevation angle at the center frequencies of L5 band and L1 band as shown in Figure 8 and 9, respectively. Linear average gain is the antenna gain (dBi) averaged in linear form over all azimuth (0-360˚) angles, axial ratio (dB) is using the same way to calculate.

In the Figure 8 (a), along the horizontal axis, the elevation angles from 0° to 90° are divided into increments of 6º. The dashed curve represents the average axial ratio in unit decibel (dB), whose values are read from the left vertical axis. The solid curve represents the LAG in dBi results, whose values are read from the right vertical axis. The legend [Simulation EVB1176] stands for the standalone patch antenna on a 250 mm circular ground plane at 1176 MHz (Figure $2(b)$) of the EM simulation result; the legend [EVB 1176] an evaluation board standalone patch antenna on a 250mm circular ground plane at 1176 MHz of indoor antenna range (Figure $3(a)$); the legend [Sharkfin 1176] stands for the compacted multi-function shark-fin antenna on a 250mm circular ground plane at 1176 MHz of indoor antenna range (Figure $3(b)$). The rest can be done in the same manner.

For the LAG, simulation results correlate well with the real antenna sample in both bands. The LAG ranges from -6 dBi to +5dBi over elevation 0° to 90°. The real-world measurements do not deviate from the simulation results by more than 2dBi within the tested range.

For the axial ratio, the discussion is separated by frequency group. In the low frequency L5 band, Figure 8 $(a)$ the simulation shows the ideal AR trend continuously decreases from horizontal (Elevation = 0˚) 10.3 dB to zenith (Elevation = 90˚) +1.1 dB. The evaluation board result has a similar trend, a better AR at lower elevation and a quite stable value at around 3 dB over the higher elevations. The indoor antenna range has reflection and ground heat noise that may be causing deviation from the simulation and EVB result. When the same antenna is assembled with other antenna systems (SDARS and telephone elements), the AR experiences degradation (above 3 dBi) at most elevation below 60°. At the higher elevations, the simulation, EVB and shark-fin all perform at a similar level.

In the Figure 8 $(b)$, it is a polar plot of radiation pattern gain at elevation 36° cut over 360° azimuth angel. The color curves represent the same antenna configuration as the Figure 8 $(a)$. For the EVB sample, both simulation (black curve) and real antenna (blue curve) are matched, the gain on the azimuth $\varphi$ direction has omni-directional the radiation pattern. The axial ratio is kept under 4 dB. When the GNSS antenna integrated into the shark-fin packaging (red curve), due to the previous discussion on the boundary condition Equation $(6a - 6c)$ and simulation result on Figure 6. The horizonal plane gain become directivity and the AR is kept at level of 6 dB.

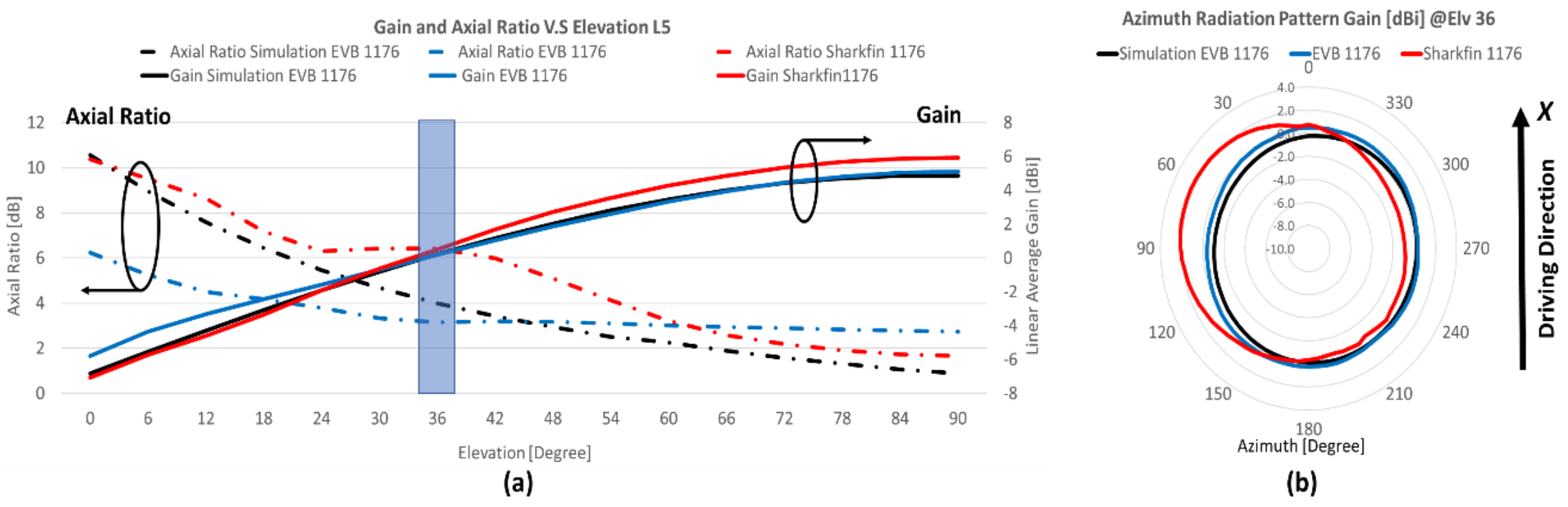

Fig. 8. (a) Linear Average Gain and Axial Ratio over Elevation at L5 Band; (b) L5 Band Azimuth Gain @ Elv 36 degree

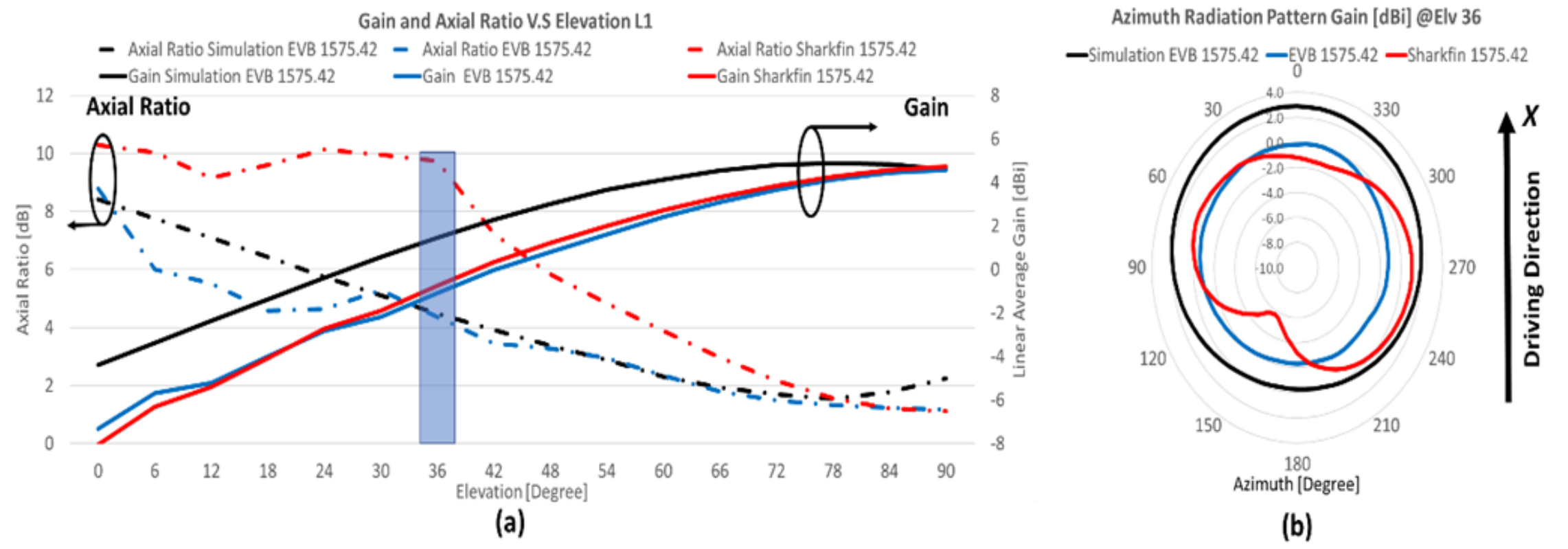

Fig. 9. (a) Linear Average Gain and Axial Ratio over Elevation at L1 Band; (b) L1 Band Azimuth Gain @ Elv 36 degree

In the higher frequency L1 band in Figure 9 $(a)$, the simulation and EVB results are also quite matched and the shark-fin results exhibit a similar performance degradation at lower elevation angels ($< 36°$) as well. It is clearly to show on the right side of Figure 9 $(b)$, the shark-fin GNSS antenna omni-directivity has further degradation, especially on the rear orientation. Recalling the simulation result of current density on Figure 6. The GNSS L1 band would have influence from vertical Telephone1 element's coupling surface current due to the cellular resonate frequency Band 3 nearby. Due to this resonate current coupling, the patch antenna current densities get disturbed and antenna radiation pattern has a null on the rear direction. Thus, the circularly polarized fields get degradation, it is the reason the axial ratio on the left side is 9.7 dB on elevation 36°.

The real measurement results show that when the GNSS antenna is assembled with other elements in a compact multifunctional package, the electrical performance is affected.The simulation result shows the antenna in a perfect electrical wall boundary condition Equation $(6a)$ and $(6b)$, therefore the antenna performance is nearly the theory calculation with a good agreement in circular polarization, such as axial ratio. In the meanwhile, the two telephone antennas could be treated as directors or reflectors for the patch antenna, that will break the perfect magnetic wall boundary condition Equation $(6c)$, therefore, the shark-fin GNSS antenna field radiation pattern and phase performance will have a deviation with simulation and EVB antenna's result. The SDARS antenna has a similar ceramic construction, albeit a smaller physical size. The result shown is the best tuning of each antenna and the physical distance between each element is maximized. Regardless, interference between each of the antenna elements in the co-existing system is inevitable. The minimize the influence and prioritize key performance metrics within a desired range, in this case the target level is an AR below 3dB above 60° elevation.

## 8. OUTDOOR ANTENNA RANGE SETUP

After the standalone patch and multi-function shark-fin antenna are tested in the indoor lab, on-vehicle measurement is made at Oakland University's quasi-far-field antenna range, shown in Figure 10. The shark-fin antenna is mounted on a sedan vehicle with a curved roof, meaning the shark-fin antenna has a proper ground connection on the top of the car's roof and testing accurately portrays real world performance on a curved ground plane [12-13]. A zoomed-in detail of the mounted antenna is provided at the left bottom corner. This vehicle is parked on an electronically controlled rotating steel turntable (electrical ground plane). The transmitter is a cross-dipole parabolic reflector antenna (same as indoor transmission antenna, Figure $7c$) that is mounted on a white fiberglass arched arm, which can adjust the height of the arm's rotational axis in order to match with the vehicle height. The purpose of this kind of design is to enable the parking of the vehicle, centering the AUT on the turntable. The transmission antenna can illuminate the car roof in a spherical coordinate system. A comparison of the post-processed data from the indoor and outdoor ranges is provided in the next section.

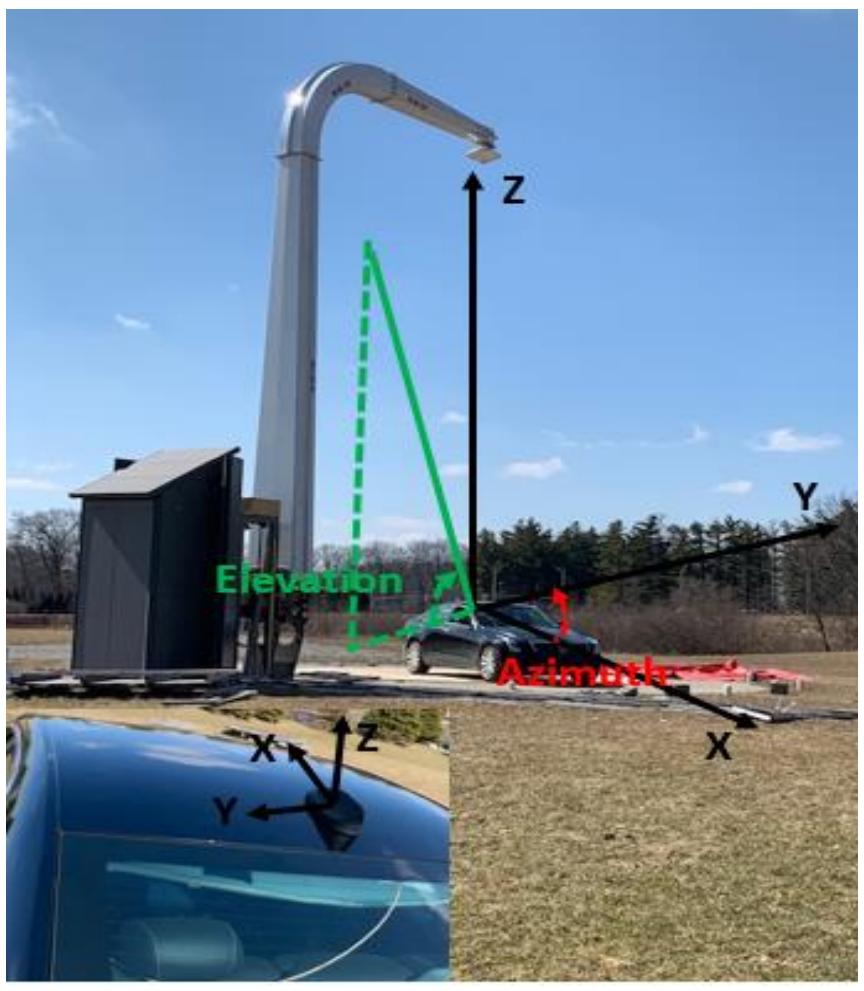

Fig. 10. Oakland University Outdoor Antenna Range

## 9. PHASE CENTER VARIANT AND PHASE CENTER OFFSET

In antenna design theory, the phase center is defined as *the point from which the electromagnetic radiation spreads spherically outward, with the phase of the signal being equal at any point on the sphere* as demonstrated in Figure 11. If the antenna phase contour is ideally spherical, the phase center is the physical center point of the antenna, also called the antenna reference point (ARP) [14-15]. However, in a real situation, the antenna phase contour cannot be ideally spherical. The real far-field phase deviates away from the ideal phase, dependent on signal directionality. This deviation away from the mean ideal phase is called PCV. A lower PCV corresponds to a stable phase variant $\varepsilon_{r,PCV}$ in Equation $(4)$. Measurements of the antenna's received phase can be used to calculate back its phase center. For each elevation and arc length, the theoretical origin of the electromagnetic wave in that specific direction can be calculated from the real phase of the measured signal. This is done by creating a triangle whose two outer vertices are defined by the angles of two adjacent measurement points and their measured phases and whose inner vertex (the real phase center) is defined as the calculated signal origin of that would produce the phase measured at these points. *The distance from this signal origin to the ARP* is the PCO. The initial correction of the PCO can only be done provided the distance to the transmitted signal is accurately known. A lower PCO represents a concentrated phase center of the AUT. So PCV $\varepsilon_{r,PCV}$ and PCO $\varepsilon_{r,PCO}$ are importance parameters in evaluating the antenna phase performance in Equation (4).

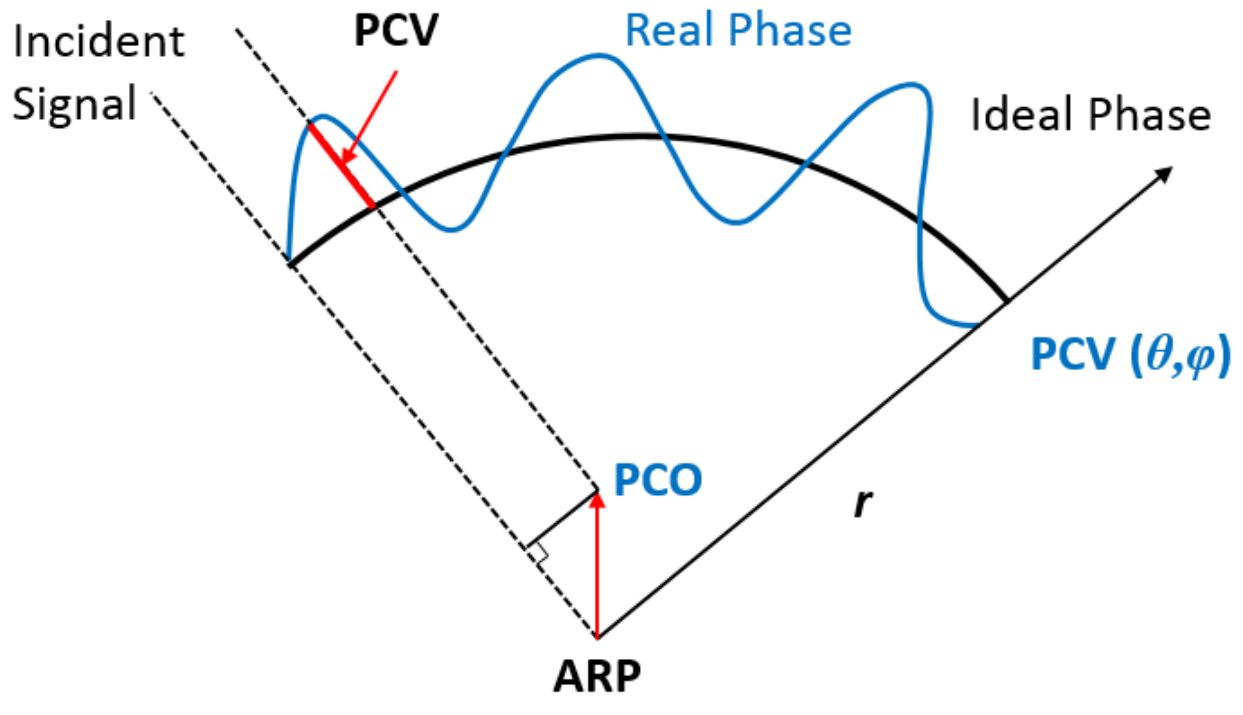

Fig.11. PCV and PCO Model (as per Zeimetz and Kuhlmann, 2006 [16])

Furthermore, varying frequency will cause changes in phase center due to the difference in wavelengths of Equation $(5a)$ and $(5b)$. The phase center determination of this paper is presented at the center frequency of the L5 band (1176 MHz) and the L1 band (1575.42 MHz) in spherical phase method based on a least squares method [16], which requires an upper hemisphere far-field radiation pattern. It is including phase information $Phase\ (\theta, \varphi)$, following the spherical coordinate conventions. A far-field phase expansion can be expressed as

$$Phase\ (\theta, \varphi) = \frac{2\pi}{\lambda}(x \cdot sin\theta cos\varphi + y \cdot sin\theta sin\varphi + z \cdot cos\theta) \qquad (7)$$

Let the phase center origin be located at PCO$(x, y, z)$ , be the unknown parameter to be solved and $\lambda$ is the wavelength of the measured signal in millimeters [mm].
Rewriting in matrix from (7) gives

$$Phase\ (\theta,\varphi) = \frac{2\pi}{\lambda}[sin\theta cos\varphi \quad sin\theta sin\varphi \quad cos\theta] \cdot \begin{bmatrix} x \\ y \\ z \end{bmatrix} \tag{8}$$

$$\begin{bmatrix} x \\ y \\ z \end{bmatrix} = \frac{\lambda}{2\pi}[sin\theta cos\varphi \quad sin\theta sin\varphi \quad cos\theta]^{-1} \cdot Phase\ (\theta,\varphi) \tag{9}$$

The antenna's received phase from different angle of arrivals (AOA) is extracted from the far-field measurement results from the calculation. If an AOA phase correction could adjust the preliminary phase measurement result in advance, then the localization system could recall the predefined PCV result to calculate the distance. That could solve antenna phase-center variation error $\varepsilon_{r,PCV}$ in Equation (4) within the last one period [7], which would improve the carrier phase accuracy.

The range measures the received signal $Phase\ (\theta,\varphi)$ and using this phase data, the PCV could be calculated by the wavelength. Furthermore, the final PCV result for a given angle can be represented in millimeters [mm].

$$PCV = \frac{Phase\ (degree)}{360^{\circ}} \times \frac{c\ (speed\ of\ light)}{Frequence\ (Hz)} [mm] \tag{10}$$

Figure 12 $(a)$ shows the PCV $Phase\ (\theta,\varphi)$, at the center of the L5 band (1176 MHz) on an evaluation board, which presents in Figure 3 $(a)$. In the PCV $Phase\ (\theta,\varphi)$'s result, the horizontal axes are elevation and azimuth angles while the vertical axis is PCV shown as a Jet colormap to represent the variation in millimeter. At zenith (Elevation = 90°), the phase is zero, and this is used as the phase pattern ARP. As elevation decreases, the PCV changes from -36.3 to +16.5 mm over elevation and azimuth.

When the patch is placed in a shark-fin antenna as illustrated in Figure 3 $(b)$, the PCV's result increases as shown in Figure 12 $(b)$, the total PCV range enlarges to -57.7 to +24.4 mm.

After the indoor range testing, the outdoor range is used to test the antenna on-vehicle at Oakland University (Figure 10). The same shark-fin antenna is placed on a test vehicle and the result is represented in Figure 12 $(c)$. The PCV result shows a similar trend as discussed for the indoor EVB and Shark-fin. However, the variation increases from -86.7 to +10.7 mm.

Moreover, the maximum phase variation occurs along the horizontal plane (Elevation = 0°) when the antenna is directed toward the front direction (Azimuth = 0°) and when the antenna is directed toward the rear direction (Azimuth =180°). It is clear to see that the compact shark-fin structure and on-vehicle scenarios have a large influence on the antenna phase performance. The theory behind the phenomenon is when the antenna mounted on the roof, the ground connection is not a perfect electrical wall boundary condition Equation $(6a)$ and $(6b)$, the patch electrical field was destroyed by the repulsive mechanism of the bottom of the patch, thus the charge concentration will not dominate by the top layer attractive mechanism [17]. As the similar explanation in Section V, the antenna phase performance degradation due to the field distribution. A closed form expression of a finite ground and imperfect boundary condition is hardly to be found in a mathematics work. The EM software could provide a simulation result as a guideline. In order to evaluate a high precision automotive GNSS antenna phase center performance properly, a real vehicle testing is essential during the design work.

Figure 13 $(a)$, 13$(b)$, and 13$(c)$ show the center frequency of L1 band (1575.42 MHz) PCV results for the GNSS dual-band patch in an EVB on a 250mm circular ground plane, in a shark-fin antenna on a 250mm circular ground plane and in a shark-fin antenna on a vehicle-roof, respectively. The conclusion is almost identical as for the L5 frequency. The EVB antenna has the best PCV performance, while the on-vehicle testing points to the real word scenario has the worst PCV.

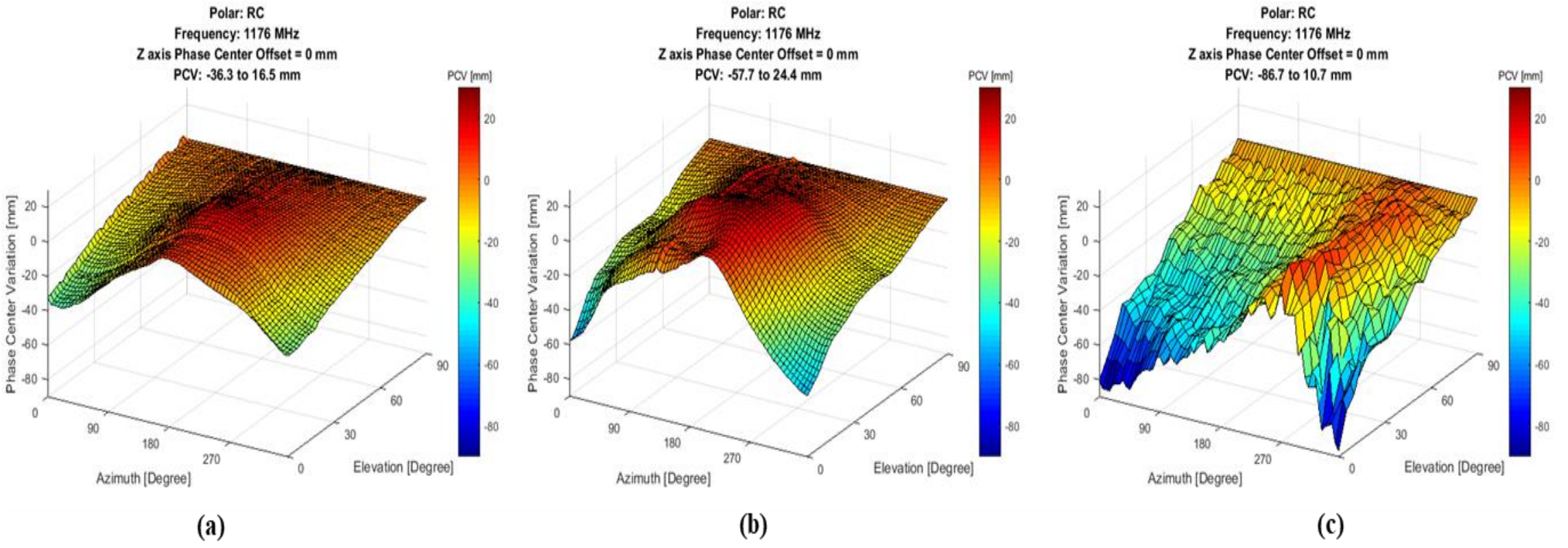

Fig. 12.
(a) PCV at 1176MHz **@EVB**;
(b) PCV at 1176MHz **@Shark-fin**;
(c) PCV at 1176MHz **@On-vehicle**

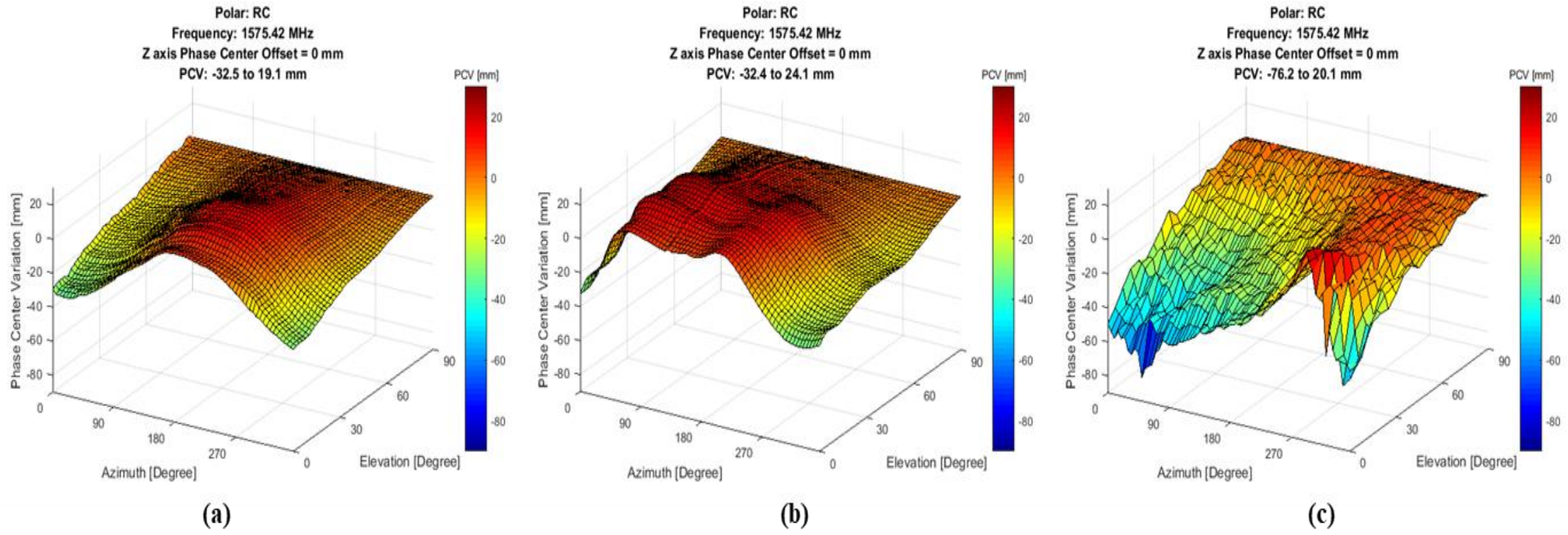

Fig. 13.
(a) PCV at 1575.42MHz @**EVB**;
(b) PCV at 1575.42MHz **@Shark-fin**;
(c) PCV at 1575.42MHz **@On-vehicle**

Up to now, the $Phase\ (\theta, \varphi)$ determination has been from a far-field measured at indoor range (Figure 12 $(a)$, 12 $(b)$, 13 $(a)$) and 13 $(b)$) and outdoor range (Figure 12 $(c)$ and 13 $(c)$). In order to find the PC$O(x, y, z)$, the calculation could be done by pseudo-inverse Equation (11).

$$PCO\begin{bmatrix}x\\y\\z\end{bmatrix} = \frac{\lambda}{2\pi}\cdot\begin{bmatrix} sin\theta_1 cos\varphi_1 & sin\theta_1 sin\varphi_1 & cos\theta_1 \\ sin\theta_1 cos\varphi_2 & sin\theta_1 sin\varphi_2 & cos\theta_1 \\ \vdots & \vdots & \vdots \\ sin\theta_1 cos\varphi_N & sin\theta_1 sin\varphi_N & cos\theta_1 \\ sin\theta_2 cos\varphi_1 & sin\theta_2 sin\varphi_1 & cos\theta_2 \\ sin\theta_2 cos\varphi_2 & sin\theta_2 sin\varphi_2 & cos\theta_2 \\ \vdots & \vdots & \vdots \\ sin\theta_M cos\varphi_{N-1} & sin\theta_M sin\varphi_{N-1} & cos\theta_M \\ sin\theta_M cos\varphi_N & sin\theta_M sin\varphi_N & cos\theta_M \end{bmatrix}^{-1} \cdot \begin{bmatrix} Phase\ (\theta_1,\varphi_1) \\ Phase\ (\theta_1,\varphi_2) \\ \vdots \\ Phase\ (\theta_1,\varphi_N) \\ Phase\ (\theta_2,\varphi_1) \\ Phase\ (\theta_2,\varphi_2) \\ \vdots \\ Phase\ (\theta_M,\varphi_{N-1}) \\ Phase\ (\theta_M,\varphi_N) \end{bmatrix}$$

$$where: 0 \le \theta \le 90°, \quad 0 \le \varphi \le 360° \tag{11}$$

PCO $(x, y, z)$ is represented as each point corresponding to an averaged azimuth ($\varphi_N$: $0 \le \varphi \le 360°$) over each elevation level $\text{PCO}_{\theta_M}(x, y, z)$. As one of example in EVB at L5 1176 MHz Figure 12 $(a)$, it shows the PCO result in 3D view in Figure 14. In this figure, the blue dots are PCO $(x, y, z)$ in different elevations. Most of points are concentrated around the mean value, but at low elevation position's results are separated from the others. The reason is that when the antenna receives the incoming signal far below cut-off Brewster Angle, the patch antenna is unable to receive the majority of the horizontal component of the circularly polarized wave, thus appearing to be received as a vertically polarized wave. It is another proven that a poor PCO correlated with high axial ratio (as shown in Figure 8 and 9). Moreover, the gain is low at these low elevations and the AR is not good enough to accurately distinguish the correct polarization. In the post-processing, low elevation data would be taken out from GNSS receiver, thus is well known the elevation mask. The rest of test scenarios could analysis in the same manner.

In automotive navigation, the focus is on the *XY* plane, also known as the horizontal plane. Figure 15 (a) result is shown only on the *XY* cut view from the 3-dimentional result from Figure 14. A common way to evaluate the PCO precision is via circular error probable (CEP). *CEP50 is defined as the radius of a circle, centered on the mean, whose boundary is expected to include the landing points of 50% of the rounds*. CEP68 (1σ) and CEP95 (2σ) have a similar definition. In the same example EVB at L5 1176 MHz result 2D view in Figure 15 (a), CEP50 is 1.9 mm. The meaning behind this number is that without any phase correction, 50% of incident waves' phase centers will be in a radius of 1.9 mm range on *XY* plane (horizonal plane). There was similar result of 68% and 95% of waves' phase centers are in 2.3 mm and 3.5 mm range.

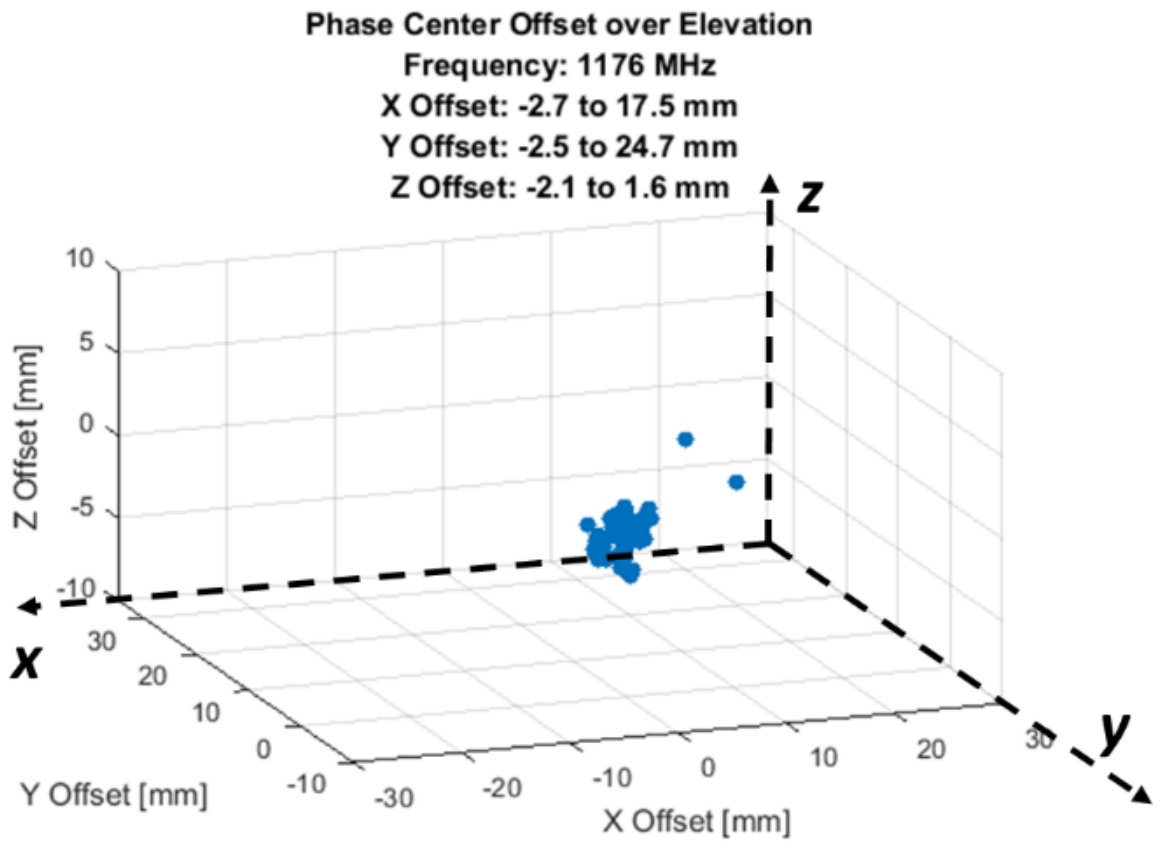

Fig. 14. 3D view PCO at 1176MHz @**EVB**

By using PCV shark-fin result from Figure 12 $(b)$, a similar 2D $XY$ plane result could be calculated in Figure 15 $(b)$, which has more PCO CEP50 (4.4 mm), CEP68 (5.4 mm) and CEP95 (8.3mm).

The PCO resulting from the shark-fin antenna on-vehicle testing are presented in Figure 15 $(c)$. The on-vehicle result shows a degradation (CEP50: 5.1mm) as compared to a standalone antenna on an EVB (CEP50: 1.9mm).

A similar result is found for the L1 frequency as shown in Figure 16 $(a-c)$ in the same three test scenarios. PCV range and PCO's CEP summarized results as tabulated in the Table 3.

In general, the same patch antenna has the best performance of PCV and PCO on a 250mm circular ground plane EVB. Whenthe patch antenna integrated in a compact shark-fin housing; the antenna phase performance is degraded. The antenna phase performance degrades further as the shark-fin antenna is installed on the car roof. The reason of this degradation is when the antenna ground connection changed, the patch antenna boundary condition changed as well. The patch antenna is grounded via the underneath PCB and shark-fin antenna chassis. This mechanical chassis is mounted on the roof by a fastener.

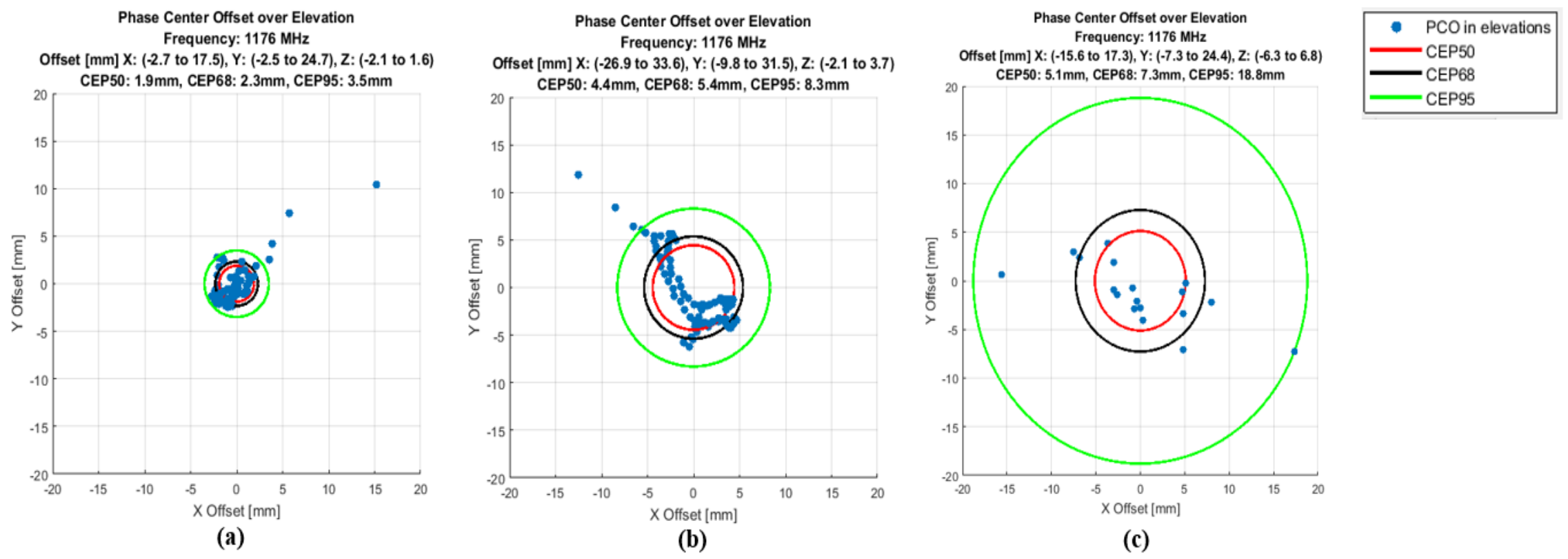

Fig. 15.
(a) 2D PCO at 1176MHz**@EVB**;
(b) 2D PCO at 1176MHz **@Shark-fin**;
(c) 2D PCO at 1176MHz **@On-vehicle**

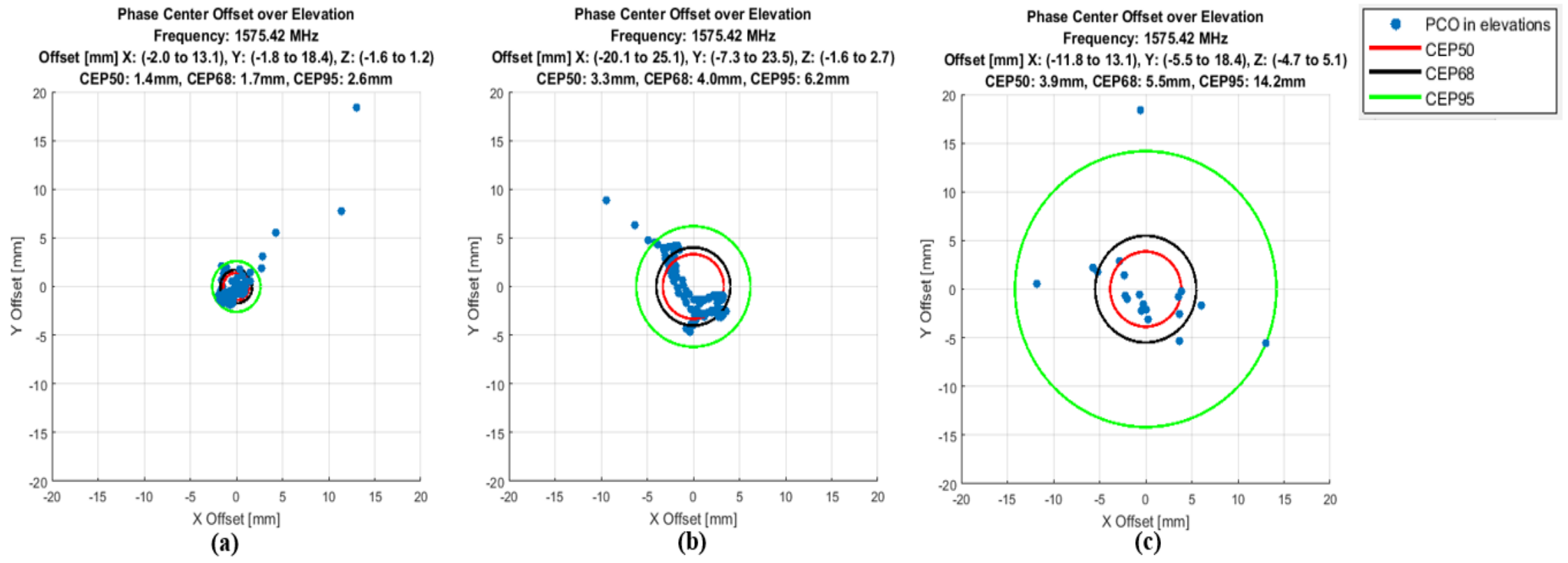

Fig. 16.
(a) 2D PCO at 1575.42MHz **@EVB**;
(b) 2D PCO at 1575.42MHz **@Shark-fin**;
(c) 2D PCO at 1575.42MHz **@On-vehicle**

Due to the limited size of this fastener metal connection, the patch antenna electrical wall, Equation ($6a$ $and$ $6b$) could be disturbed by the shark-fin antenna installation. Inside the antenna radome, the more complex element surrounding breaks the patch antenna slot's magnetic current density and resulting antenna electrical field, which in turn impacts the antenna's gain, axial ratio and phase performance. As a conclusion that, the antenna's performance has a strong connection with its surrounding environment.

Table 3. PCV and PCO' CEP

| L5: 1176MHz | | | |
|---|---|---|---|
| | **EVB** | **Shark-fin** | **On-vehicle** |
| PCV [mm] | -36.3 to 16.5 | -57.7 to 24.4 | -86.7 to 10.7 |
| PCO | | | |
| CEP50[mm] | 1.9 | 4.4 | 5.1 |
| CEP68[mm] | 2.3 | 5.4 | 7.3 |
| CEP95[mm] | 3.5 | 8.3 | 18.8 |
| L1: 1575.42MHz | | | |
| PCV [mm] | -32.5 to 19.1 | -32.4 to 24.1 | -76.2 to 20.1 |
| PCO | | | |
| CEP50[mm] | 1.4 | 3.3 | 3.9 |
| CEP68[mm] | 1.7 | 4.0 | 5.5 |
| CEP95[mm] | 2.6 | 6.2 | 14.2 |

## 10. CONCLUSION AND FUTURE WORK

In high accuracy positioning, single, double and triple differenced observables could eliminate the major error sources, except antenna phase error. This paper presents a method to solve phase error $\varepsilon_{r,PCV}$ and $\varepsilon_{r,PCO}$ in Equation (4) in automotive applications. From a classic method to evaluate the antenna by using gain and axial ratio, moreover by using the far-field phase data to show the PCV and calculated PCO based on the PCV result. Although gain and axial ratio results do not contain the phase information, they have a strong correlation with PCV and PCO result. A decent gain and axial ratio imply a decent phase result as well. They provide an alternative way to evaluate PCV and PCO quickly at early of the development stage. As the requirement of precise GNSS navigation evolves, it requires PCV measurement and PCO calculation at the validation stage.

The work presently focused on a low-cost dual-band L1/L5 GNSS patch antenna to characterize its PCV and PCO in a standalone patch antenna and in a shark-fin multi-function antenna vehicle level at indoor and outdoor measurement ranges. Due to the shark-fin antenna's complex interior configuration, such as nearby radiator, which detunes the GNSS patch antenna electrical wall and magnetic wall boundary condition. By the impact of that, the GNSS antenna's PCV and PCO results exhibited worse performance than the standalone antenna on a ground plane. A proper solution is to fix phase-center motion error $\varepsilon_{r,PCV}$ and $\varepsilon_{r,PCO}$ in Equation (4) on high precise positioning application. Each unique vehicle platform and antenna location requires the calibration profile of PCV and PCO. By using this antenna calibration file (.atx) could estimate the antenna phase-center motion error-free result.

However, this study is not yet completed. Future work includes:

1. A verification of the research methodology in multiple vehicles and multiple locations.
2. Incorporating the PCV and PCO data with real GNSS receiver carrier phase observables to verify improvement in position precision and accuracy.

## ACKNOWLEDGEMENTS

The authors would like to thank Continental Advanced Antenna Inc, North America, Inc. for the use of their indoor range. Great appreciation is also given to Mr. Andreas Fuchs, Mr. Ehab Rahman, Mr. Ryan Dombrowski and Mr. Thomas Chan for their insight.